\def\gsim{\;\lower4pt\hbox{${\buildrel\displaystyle >\over\sim}$}\,}
\def\lsim{\;\lower4pt\hbox{${\buildrel\displaystyle <\over\sim}$}\,}

\documentclass[usenatbib]{mn2e}
\usepackage{graphicx}

\newcommand\E[1]{\times10^{#1}}
\newcommand\un[1]{{\,\rm #1}}
\newcommand\rs[1]{_\mathrm{#1}}
\newcommand\g{$\gamma$}
\newcommand\ApJ{ApJ}

\title[MF and images of SN~1006]{Constraints on magnetic field strength in the remnant SN~1006 from its nonthermal images}
\author[Petruk O. et al.]{O.~Petruk$^{1,2}$, T.~Kuzyo$^{2}$, F.~Bocchino$^{3}$\\ 
$^{1}$Institute for Applied Problems in Mechanics and Mathematics, Naukova St.\ 3-b,
   79060 Lviv, Ukraine\\
$^{2}$Astronomical Observatory, National University, Kyryla and Methodia St.\ 8, 79008 Lviv, Ukraine\\
$^{3}$INAF — Osservatorio Astronomico di Palermo, Piazza del Parlamento 1, 90134 Palermo, Italy\\
}

\begin{document}

\date{Accepted .... Received ...; in original form ...}

\pagerange{\pageref{firstpage}--\pageref{lastpage}} \pubyear{2008}

\maketitle

\label{firstpage}

\begin{abstract}
Images of SN~1006 have a number of important properties. For instance, 
the bright limbs coincide spatially in various bands, they have different 
brightness, and the contrast of brightness varies from radio to gamma-rays. 
The reasons of such properties and the role of the magnetic field strength 
are discussed. Simple, almost model-independent methods and analytical 
approximations for the derivation of the strength of magnetic field 
from the comparison of radio, X-rays and TeV images of SNR are presented. 
The methods require the TeV image to be well resolved and accurate, 
at least to the level of the radio and X-ray maps, in 
order to put reasonable constraints on magnetic field. 
If we apply it to the present HESS data, they limit the strength of 
magnetic field in limbs of SN~1006 to values lower than few hundreds micro Gauss. 
If applied to the Fermi-LAT band, the model predicts
same position and same ratio of the surface brightness for GeV photons 
as for the radio band. We conclude that TeV and GeV future
high-resolution data may be very informative about the MF of SN~1006.
\end{abstract}


\section{Introduction}

Supernova remnant (SNR) SN~1006 is one of the most interesting objects for studies of Galactic cosmic rays. 
It is quite symmetrical with rather simple bilateral morphology in radio \citep[e.g.][]{pet-SN1006mf}, nonthermal X-rays \citep[e.g][]{SN1006Marco} and TeV \g-rays \citep{HESS-SN1006-2010}. Its prominent feature is the positional coincidence of the two bright nonthermal limbs in all these
bands, including TeV \g-rays. 

Could such coincidence  be sign for leptonic origin of the TeV emission? 
Detailed numerical modeling reveals factors determining the formation of the patterns 
of radio, X-ray and \g-ray maps of SNRs \citep{pet-SN1006mf,thetak,xmaps}.
TeV $\gamma$-ray emissivity does not directly depend on the magnetic field (MF) strength $B$. 
However, in the regions with 
the larger post-shock magnetic field, the radiative losses of 
relativistic electrons are higher. Therefore, their distribution downstream of 
the shock have to be steeper and the IC emission should be lower. 
In contrast, X-ray brightness 
increases in regions with higher MF because emissivity is proportional to $B^{3/2}$. 
Thus, the coincidence of the TeV \g-limbs with X-ray and radio limbs may not be directly considered 
as confirmation of the leptonic nature of \g-rays in SNR. 
Nevertheless, the azimuthal variation of brightness depends not only on the azimuthal variation of MF, 
but also on the azimuthal dependences of the injection efficiency (fraction of density of accelerated particles) 
and of their maximum energy, 
that, under some configurations, can provide correlation of the non-thermal brightness in different bands 
\citep{thetak,xmaps}. 

In order to check whether the observed correlation of the \g-ray and X-ray brightness may be considered as argument 
in favor of the IC mechanism of TeV \g-ray emission in SN~1006, a model-independent approach is developed in \citet{Petetal09icp}. 
It bases on the idea that the radio and hard X-ray data contain information about 
variations of the fraction of density and of the maximum energy of accelerated electrons. 
The radio and X-ray observations, used with a number of reliable assumptions about MF distribution,  
are used in \citet{Petetal09icp} to generate image of SN~1006 it should have in TeV \g-rays and to compare it with 
observations. 
It is shown that the pattern of the IC image of SN~1006 is naturally similar to its radio and X-ray maps, 
in various MF configurations. This is a new argument in favor of the leptonic origin of 
\g-rays from SN~1006. 

The further important property of SN~1006 is that the two bright limbs have different brightness. 
In addition, the ratio of brightness of the limbs is different in radio, hard X-ray and TeV \g-ray bands. 
In the present note, we demonstrate that location of the \g-ray limbs and the contrast of brightness 
in various bands depend on the magnetic field strength. 
Thus, it may be used to put constraints on MF. 

We exploit the approach of \citet{Petetal09icp} in an inverse manner. 
Namely, once the TeV \g-ray map of SN~1006 is known, 
we use, beside radio and hard X-ray data, properties of the TeV \g-ray image to put constraints on MF. 

Namely, in Sect.~\ref{sn1006cp:a_new_method_for_B}, we present a new simple method to estimate the strength of MF in SNR 
from contrasts of brightness in nonthermal limbs and apply it to SN~1006. 
Another realization of the method to constrain MF, from location of the TeV \g-ray limbs, is proposed in Sect.~\ref{sn1006gmf:sect_gamma_inversion}. 
Application of the methods to the GeV \g-ray image is discussed in Sect.~\ref{sn1006gmf:GeVimage_section}.
Sect.~\ref{sn1006gmf:conclusion_section} concludes. 

Our approach is model-independent in the sense that it is independent from details of CR acceleration 
and MHD, it does not involve simulations based on theoretical principles, it is free of any assumptions about SNR energy, distance, structure of interstellar medium and interstellar magnetic field etc. The only assumptions we make are the shape of the electron energy spectrum (which is rather general) and that TeV \g-ray emission is dominated by the inverse-Compton process.

\section{Magnetic field strength from 
the relative brightness of the radio, X-ray and \g-ray limbs}
\label{sn1006cp:a_new_method_for_B}

The ratio of brightness between the two bright limbs in SN~1006 varies with frequency. In particular, the NE limb is brighter in X-rays while the SW limb is more prominent in radio \citep[see Figs.~1, 2 or 11 in][]{Petetal09icp}. If the \g-ray emission of SN~1006 is dominated by IC process, then the ratio of the \g-ray brightness of these limbs depends on the strength of MF in SNR. 


Let us consider the uniform plasma with the electron distribution consisting of the (variable) power-law plus the (parameterized) exponential cut off 
\begin{equation}
 N(E)=KE^{-s+\delta s(E)}\exp\left[-\left(E/E\rs{max}\right)^{\alpha}\right]
 \label{ICpred:Espectrum}
\end{equation}
where $s$ is the spectral index, $K$ the normalization, $E\rs{max}$ the maximum energy, $\alpha$ the broadening parameter. The function $\delta s(E)$ reflects possible concavity of the spectrum as result of the efficient acceleration \citep[e.g.][]{Ber-Ell-simple_model,Blasi2002}; it increases monotonically from $0$ at around $E\simeq m\rs{e}c^2=0.5\un{MeV}$ to $\leq 0.5$ for $E\simeq E\rs{max}\sim 10\un{TeV}$ \citep[e.g.][]{Ber-Ell-simple_model,kang-et-al-2009}. This function produces negligible changes to our results, since it varies very slowly over decades in energy (for details see Appendix \ref{sn1006gmf:appendix}). Therefore, we use $\delta s=0$ hereafter. 

Parameter $\alpha$ regulates the broadening/narrowing of the high-energy end of the electron spectrum. 
Different models predicts different values for the parameter. In a number of theoretical models the cut off is broader than the pure exponent \citep[see references in][]{Reyn-Keoh-1999}, suggesting $\alpha<1$. Such broadening seems to be observed in few SNRs, including SN~1006 \citep{Ell-et-al-2000,Ell-et-al-2001,Lazendic-et-al-2004,Uchiyama-et-al-2003}. 
It should be attributed to the physics of acceleration \citep{Pet06}, 
rather than to the effect of superposition of spectra in different conditions 
along the line of sight \citep{Reyn96}. 
There are also some recent theoretical evidences that 
the cut-off may be narrower than the pure exponent depending on how strongly  
the diffusion coefficient depends on the particle momentum \citep{Blasi-2010}.
In particular, the value $\alpha=2$ appears in number of numerical and analytical approaches 
\citep{Zirak-Ahar-2007,Blasi-2010,Kang-Ryu-2010,kang-2011} with arguments that such value is relevant to the Bohm diffusion. 
The time evolution might decrease this $\alpha$ to a value between 1 and 2 \citep{Schure-et-al-2010}.
In the approximations developed below, we allow $\alpha$ to vary from 0.5 to 2.

Synchrotron radio and IC \g-ray (at 1 TeV) emissivities of electrons destributed with Eq.~(\ref{ICpred:Espectrum}) 
are approximately related as 
\begin{equation}
 q\rs{ic}\propto q\rs{r} B^{-(s+1)/2}
 \exp\left(-\beta\rs{ic}\beta\rs{\alpha ic}\left(\frac{c\rs{1}B}{\nu\rs{break}}\right)^{0.375\beta\rs{\alpha x}}\right).
 \label{ICpred:eq7}
\end{equation}
where $\nu\rs{break}$ 
is a critical frequency of the synchrotron radiation of electrons with energy $E\rs{max}$ in magnetic field with strength $B$,  
$c\rs{1}=6.26\E{18}\un{cgs}$, 
\begin{equation}
\begin{array}{l}
 \beta\rs{ic}=\left\{
  \begin{array}{ll}
   15,        &\ \mathrm{for}\ 2\leq s\leq2.5\\
   15+2(2-s), &\ \mathrm{for}\ 1.8\leq s<2\\
  \end{array} 
  \right. ,\\[12pt]
 \beta\rs{\alpha ic}=1.08\alpha^2-0.3\alpha+0.22, \\[6pt] 
 \beta\rs{\alpha x}=\alpha(1.24-0.24\alpha). 
\end{array}
\end{equation}
Details are considered in \citet{Petetal09icp} where the value $\alpha=1$ was adopted (then $\beta\rs{\alpha ic}=\beta\rs{\alpha x}\equiv1$).
The relation (\ref{ICpred:eq7}) is a generalization of Eq.~(11) in \citet{Petetal09icp}; it is valid\footnote{The maximum error of the approximation (\ref{ICpred:eq7}) is 30\% for $s$ from 1.8 to 2.5, $\alpha$ from 0.5 to 2 and values of $E\rs{max}$ which provide non-negligible contribution of the spectrum (\ref{ICpred:Espectrum}) to \g-ray emission at 1 TeV.} for any value of $\alpha$ from 0.5 to 2.

Let's 
consider two regions on SNR image\footnote{These regions have to be rather small to allow for assumption of uniform MF there.} 
with the same strength $B$ (for example, some parts of the two bright limbs in SN~1006) and denote them with 
the index `1' (NE limb) and index `2' (SW limb). 
The ratio 
${\cal R}\rs{ic}={q\rs{ic1}/q\rs{ic2}}$ 
of IC \g-ray brightness (at 1 TeV) of these regions is 
\begin{equation}
\begin{array}{lll}
 {\cal R}\rs{ic}&=&
 {\cal R}\rs{r} 
 \exp\left(-\beta\rs{ic}\beta\rs{\alpha ic}\left(c\rs{1}B\right)^{0.375\beta\rs{\alpha x}}
 \right.\\ \\ &\times& \displaystyle \left.
 \left[\frac{1}{\nu\rs{break1}^{0.375\beta\rs{\alpha x}}}-
 \frac{1}{\nu\rs{break2}^{0.375\beta\rs{\alpha x}}}\right]\right)
 \end{array}
 \label{SN1006cp:eq1}
\end{equation}
where ${\cal R}\rs{r}={q\rs{r1}/q\rs{r2}}$ is the ratio of the radio brightness.
It is apparent from this formula that the ratio of IC brightness between the two bright limbs in SN~1006 really depends on the absolute value of the magnetic field strength. 

The radio and X-ray emissivities are approximately related as 
\begin{equation}
 q\rs{x}\propto q\rs{r}
 \exp\left(-\beta\rs{x}\left(\frac{\nu\rs{x}}{\nu\rs{break}}\right)^{0.364\beta\rs{\alpha x}}\right)
 \label{ICpred:eq4}
\end{equation}
where $\beta\rs{x}=1.46+0.15(2-s)$; this relation generalizes Eq.~(8) from \citet{Petetal09icp} to $\alpha$ other than unity; its accuracy is comparable to accuracy of (\ref{ICpred:eq7}).
Therefore, the ratio of X-ray brightness between the two limbs 
${\cal R}\rs{x}={q\rs{x1}/q\rs{x2}}$ is 
\begin{equation}
 {\cal R}\rs{x}=
 {\cal R}\rs{r} 
 \exp\left(-\beta\rs{x}\nu\rs{x}^{0.364\beta\rs{\alpha x}}
 \left[\frac{1}{\nu\rs{break1}^{0.364\beta\rs{\alpha x}}}-
 \frac{1}{\nu\rs{break2}^{0.364\beta\rs{\alpha x}}}\right]\right).
 \label{SN1006cp:eq2} 
\end{equation}
If we neglect differences between the powers $0.364$ and $0.375$ assuming both equal to $0.37$, and substitute (\ref{SN1006cp:eq1}) with combination in the exponent, consisting of the break frequencies, from (\ref{SN1006cp:eq2}), we obtain that  
\begin{equation}
 {\cal R}\rs{ic}\simeq {\cal R}\rs{r} \left(\frac{{\cal R}\rs{x}}{{\cal R}\rs{r}}\right)^{\xi(B)},
 \quad \xi=\frac{\beta\rs{ic}\beta\rs{\alpha ic}(c\rs{1}B)^{0.37\beta\rs{\alpha x}}}
 {\beta\rs{x}\nu\rs{x}^{0.37\beta\rs{\alpha x}}}.
 \label{SN1006cp:eq3} 
\end{equation}
This equation relates approximately the ratios of surface brightness between the two regions with the same MF strength (e.g. the bright limbs) in radio, hard X-ray and 1 TeV \g-rays with $B$ as a free parameter. Having ${\cal R}\rs{r}$, ${\cal R}\rs{x}$ and ${\cal R}\rs{ic}$ measured in observations, one may have estimation for the value of MF in these regions. If $s=1.8\div2.2$, $\alpha=1$ and $\nu\rs{x}=2.4\un{keV}$ then $\xi=0.15
B\rs{\mu}^{0.37}$ where $B\rs{\mu}$ is in $\mu$G. 

Eq.~(\ref{SN1006cp:eq3}) shows that ${\cal R}\rs{ic}(B)$ is increasing function of $B$ and that in the limit 
$B\rightarrow 0$, the ratio ${\cal R}\rs{ic}\rightarrow {\cal R}\rs{r}$. 

\begin{figure}
\centering
\includegraphics[width=8.3truecm]{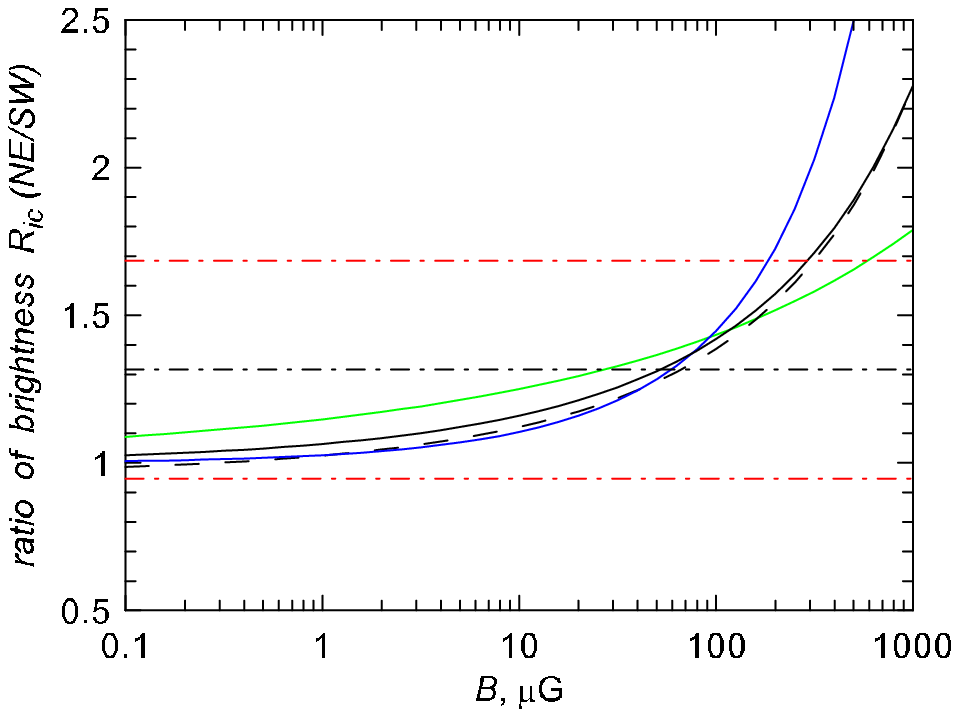}
\caption{Dependence of the ratio of IC \g-ray brightness ${\cal R}\rs{ic}$ (at 1 TeV) on $B$, as from Eq.~(\ref{SN1006cp:eq3}) and $\alpha=1$ (black lines), $\alpha=0.5$ (green line), $\alpha=2$ (blue line). The value $s=2$ is adopted; the plot is however almost insensitive to $s$.
Observed values ${\cal R}\rs{r}$, ${\cal R}\rs{x}$ used for solid lines are obtained from 
the high-resolution images; those used for dashed line are from the images smoothed to the HESS resolution. 
Horizontal lines represent the ratio of \g-ray brightness (black) and corresponding 1-$\sigma$ errors (red) from the data of \citet{HESS-SN1006-2010}. 
              } 
\label{SN1006gmf:mf-azimuth}
\end{figure}

Let's proceed to the data of observations and consider the radio (VLA+Parkes) image of SN~1006 at 1.5 MHz \citep{pet-SN1006mf} as well as its 
hard X-ray (XMM) image \citep{SN1006Marco}. These images have high resolution; for sake of comparison, we will consider also the same 
radio and X-ray images smoothed to the resolution of HESS (by the Gaussian with $2'$ sigma). 
The two regions are defined by 
an annulus ($8'$ to $20'$ from center) centered on the remnant and 
sectors $27^\mathrm{o}-45^\mathrm{o}$ from north for NE limb and $216^\mathrm{o}-234^\mathrm{o}$ for SW limb. 
Region are chosen to overlap the maximum in the azimuthal profile of \g-ray brightness and 
are $18^\mathrm{o}$ wide that corresponds to one step in the \g-ray profile reported by \citet[][their fig.~6]{HESS-SN1006-2010}. 

In SN~1006, ${\cal R}\rs{r}=0.998$ or 0.958, ${\cal R}\rs{x}=1.53$ or 1.50 for the high-resolution or the smoothed images respectively. 
The dependence of the IC brightness ratio ${\cal R}\rs{ic}$ on $B$ calculated with Eq.~(\ref{SN1006cp:eq3}) and these numbers is shown on 
Fig.~\ref{SN1006gmf:mf-azimuth}.
One can see that higher MF results in the larger difference between the brightness of the two \g-ray limbs. 
It is evident that 
lines obtained from images with the high resolution (solid lines) and from smoothed images (dashed lines) differ just a bit, because regions considered are large enough to be unaffected by the smoothing to few arc minutes. 
Fig.~\ref{SN1006gmf:mf-azimuth} reveals that the dependences ${\cal R}\rs{ic}(B)$ for various $\alpha$ are similar; though uncertainty in the value of $\alpha$ may result in a bit different estimates of $B$.

HESS report gives ${\cal R}\rs{\gamma}=1.32\pm0.37$ \citep[][Fig.~2]{HESS-SN1006-2010}. 
The later ratio together with errors is shown on Fig.~\ref{SN1006gmf:mf-azimuth} by horizontal lines. 
Unfortunately, rather slight dependence of the ratio of \g-ray brightness of the value of MF strength (few times versus variation of $B$ over decades) and the large errors in the present HESS data do not allow us to give preference to any strength of MF. Any value between few $\un{\mu G}$ and few hundred $\un{\mu G}$ falls into the 1-$\sigma$ error. 
Nevertheless, we believe that the suggested method will be useful in the future when experiments provide increased accuracy and resolution. 

\section{MF strength from location of the \g-ray limbs}
\label{sn1006gmf:sect_gamma_inversion}

In cases when MF in two regions is different, an approximate formula (\ref{SN1006cp:eq3}) may be generalized to 
\begin{equation}
\begin{array}{l}
 \displaystyle
 {\cal R}\rs{ic}\simeq {\cal R}\rs{r} \left(\frac{{\cal R}\rs{x}}{{\cal R}\rs{r}}\right)^{\xi(B_2)}
 {\cal R}\rs{B}^{-(s+1)/2}\\ \\
 \qquad\times\displaystyle
 \exp\left(-\beta\rs{ic}\beta\rs{\alpha ic}\left(\frac{c\rs{1}B_2}{\nu\rs{break1}}\right)^{0.37\beta\rs{\alpha x}}
 \left({\cal R}\rs{B}^{0.37\beta\rs{\alpha x}}-1\right)\right)
 \end{array}
 \label{SN1006gmf:eq3_contrast} 
\end{equation}
where ${\cal R}\rs{B}=B_1/B_2$. 

IC \g-ray emission is expected to be smaller where the strength of MF is higher \citep[e.g.][]{aha-ato-1999} while the synchrotron emission $\propto B^{3/2}$. This might result in an \g-ray 'limb-inverse' property for SNRs \citep{thetak}. Namely, the faintest IC \g-ray regions in SNR images might be located 
where the radio and X-ray images have maximum brightness. If this would happen in SN~1006, then IC \g-ray limbs would be expected in SE-NW regions, in contrast to radio and X-ray limbs which are at NE-SW azimuths. 

Let's now, in contrast to Sect.~\ref{sn1006cp:a_new_method_for_B}, denote with the indexes `1' and `2' the values in NE bright limb and in SE faint region of SN~1006, respectively. These two regions are separated in azimuth on $90^\mathrm{o}$, thus MF is expected to differ up to few times there; 
in particular, in the classic model of MF inside SNR (i.e when ISMF is assumed constant and the only compression of ISMF is considered which depends on the obliquity of the shock), ${\cal R}\rs{B}$ varies from 1 for an aspect angle (between the ambient MF and the line of sight) $\phi\rs{o}=0^\mathrm{o}$ to ${\cal R}\rs{B}=4$ for $\phi\rs{o}=90^\mathrm{o}$. 
Therefore, we should use Eq.~(\ref{SN1006gmf:eq3_contrast}). 
In this formula, the ratio ${\cal R}\rs{ic}$ could be higher than unity (that corresponds to the observed location of \g-ray limbs in SN~1006) or less than unity (that reflects the inversion of the \g-ray limbs). Eq.~(\ref{SN1006gmf:eq3_contrast}) shows that the value of ${\cal R}\rs{ic}$ depends, -- beside observables $s$, ${\cal R}\rs{r}$, ${\cal R}\rs{x}$ and $\nu\rs{break1}$, -- on the strength $B_2$, on contrast ${\cal R}\rs{B}$ and on $\alpha$. Thus, the observed location of \g-ray limbs can also give constraints on the MF field in SN~1006; namely, on $B_2$ and $B_1={\cal R}\rs{B}B_2$. In particular, what is the range of MF strength which would result in the inversion of the \g-ray limbs in SN~1006 and therefore may be excluded? 


\begin{figure}
\centering
\includegraphics[width=8.3truecm]{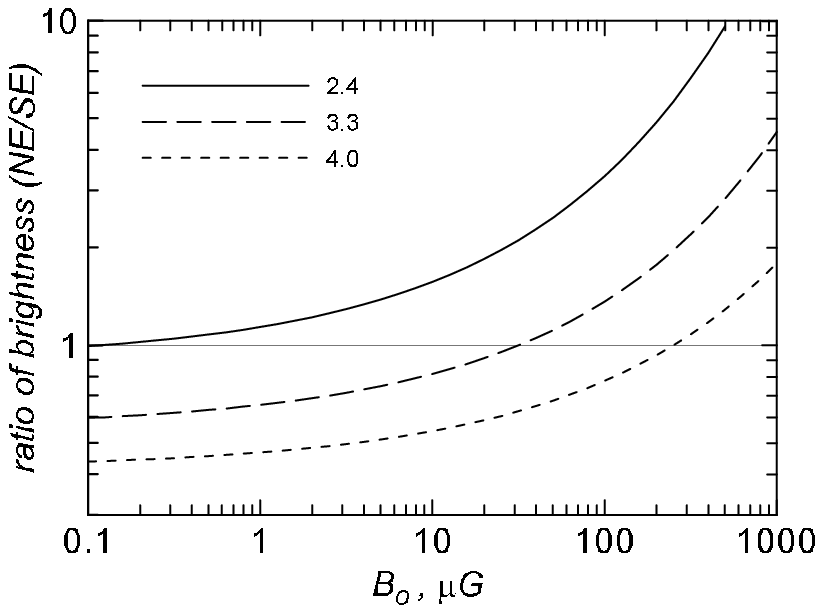}
\caption{The ratio of ${\cal R}\rs{ic}$ of IC surface brightness (at 1 TeV) in NE and SE regions of SN~1006, as it is given by Eq.~(\ref{SN1006gmf:eq3_contrast}) with observational values for ${\cal R}\rs{r}$, ${\cal R}\rs{x}$ and $\nu\rs{break1}$. 
Calculations are done for $\alpha=1$, $s=2$ and three cases of MF contrast ${\cal R}\rs{B}$, namely $2.4,\ 3.3,\ 4.0$. In the classic model of MF inside SNR  (i.e when only compression is considered which depends on the obliquity of the shock), these correspond to the aspect angles $\phi\rs{o}=70^\mathrm{o},\ 80^\mathrm{o},\ 90^\mathrm{o}$ respectively. In SN~1006, an aspect angle is $70^\mathrm{o}$ for a model of the uniform ISMF \citep{pet-SN1006mf} or $38^\mathrm{o}$ for nonuniform ISMF \citep{FB-et-al-2011}. 
${\cal R}\rs{\gamma}$ in the HESS data is $6.2\pm 5.4$ \citep{HESS-SN1006-2010}.} 
\label{sn1006:fig-IClimbs}              
\end{figure}

In the radio and hard X-ray images of SN~1006, ${\cal R}\rs{r}=3.3$, ${\cal R}\rs{x}=50$. The break frequency is measured to be $\nu\rs{break1}=9.0\E{16}\un{Hz}$ \citep{SN1006Marco}; this value corresponds to an average of the break frequencies $(5\div 20)\E{16}\un{Hz}$ measured by \citet{katsuda-2010} in a number of smaller regions. With this observables, 
Fig.~\ref{sn1006:fig-IClimbs} demonstrates, in a full agreement with estimation of \citet{Petetal09icp}, that any MF strength $B\rs{o}>1\un{\mu G}$ results in the same locations (i.e. ${\cal R}>1$) of IC, radio and X-ray limbs in SN~1006, if ${\cal R}\rs{B}<3$ and $\alpha$ is around unity. 
From physical point of view, 
larger MF in the limbs regions (i.e. ${\cal R}\rs{B}>1$, \citet{pet-SN1006mf}) results in larger radiative losses of relativistic electrons emitting TeV \g-rays. The number of such electrons have therefore to be smaller there than around SE and NW regions. 
However, in SN~1006, $E\rs{max}$ increases toward limbs \citep{sn1006cp} compensating the lack of emitting electrons due to larger MF. 

This effect works in SN~1006 for the aspect angles $\phi\rs{o}\leq70^\mathrm{o}$ or for MF contrasts ${\cal R}\rs{B}\leq 2.4$. If the aspect angle (or ${\cal R}\rs{B}$) in SN~1006 would be larger then some ranges for $B\rs{o}$ could be excluded by the observed fact that \g-ray limbs are at the same location as in the radio and X-ray images. For example, if $\phi\rs{o}=90^\mathrm{o}$ (Fig.~\ref{sn1006:fig-IClimbs} short-dashed line) then the only possible strength of ISMF could be greater than $\sim 100\un{\mu G}$. 

\begin{figure}
\centering
\includegraphics[width=8.3truecm]{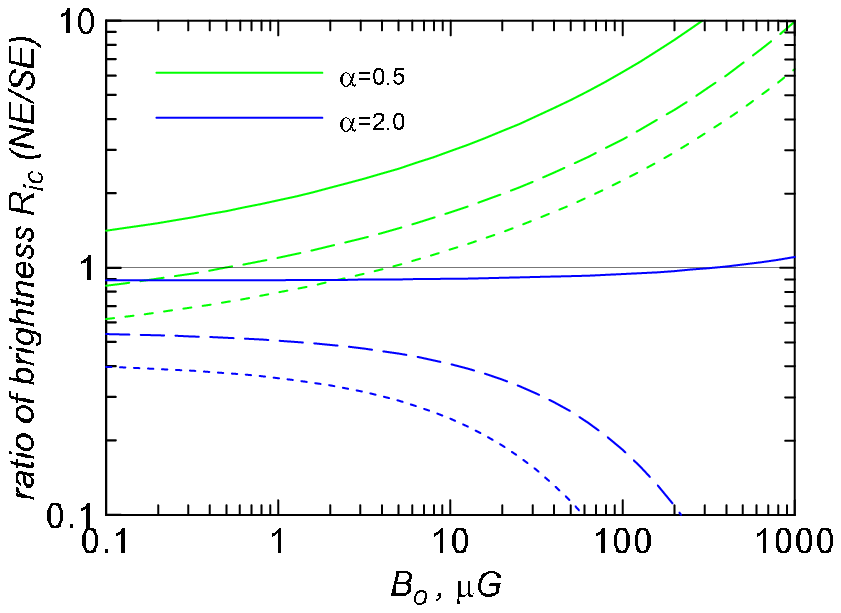}
\caption{The same as on Fig.~\ref{sn1006:fig-IClimbs} for $\alpha=0.5$ (green lines) and $\alpha=2$ (blue lines).} 
\label{sn1006:fig-IClimbs_alpha}              
\end{figure}

The role of the parameter $\alpha$, which determines the shape of the high-energy end of the electron spectrum, is demonstrated on Fig.~\ref{sn1006:fig-IClimbs_alpha}. Namely, the cut off broader than the pure exponent (e.g. $\alpha=0.5$, green lines) provides the same location of the bright \g-ray limbs even in case of larger differences of MF in two regions ${\cal R}\rs{B}$ (or larger aspect angles, up to $90^\mathrm{o}$). In contrast, the thinner end of the electron spectrum is not allowed by the observed location of the \g-ray limbs ($\alpha=2$, blue lines) and ${\cal R}\rs{B}\geq 2$ because ${\cal R}\rs{ic}<1$ for such $\alpha$.

In principle, once one has \g-ray data with high enough accuracy, namely ${\cal R}\rs{\gamma}$ between bright NE and faint SE regions, one can put limitations on MF from Fig.~\ref{sn1006:fig-IClimbs} and \ref{sn1006:fig-IClimbs_alpha}. However, present accuracy of the HESS azimuthal brightness profile prevents us from such an estimate. Really, the ratio of brightness between the NE (around azimuth $30^\mathrm{o}$) and SE (around azimuth $-60^\mathrm{o}$) rims in the HESS data \citep[Fig.~6 in][]{HESS-SN1006-2010} is $6.2\pm 5.4$.

\section{Properties of GeV \g-ray image of SN~1006 as it could be seen by Fermi telescope}
\label{sn1006gmf:GeVimage_section}

Eq.~(\ref{SN1006cp:eq3}) may be used for the \g-ray image where most of emission arises from electrons with energies 
around cut off, $E\sim E\rs{max}$. This is the case of TeV observations by HESS. 
Fermi \g-ray telescope operates in GeV \g-rays. Most of contribution to emission at these energies are from electrons with 
$E\ll E\rs{max}$. Therefore, Eq.~(\ref{ICpred:eq7}), which is written for ${\cal R}\rs{B}=1$, simplifies to $q\rs{ic}\propto q\rs{r} B^{-(s+1)/2}$ and Eq.~(\ref{SN1006cp:eq1}) to
\begin{equation}
  {\cal R}\rs{ic,Fermi}= {\cal R}\rs{r}. 
\end{equation}
Therefore, if the \g-ray emission of SN~1006 is due to the inverse-Compton process then contrast between limbs in the Fermi image should be the same as in radio. 
This will provide an important direct evidence about leptonic nature of \g-rays in SN~1006.
If contrast between limbs in Fermi image differs from the contrast in the radio image then it should be that either MF strengths in limb locations are different  or contribution to \g-rays from hadronic process is not negligible in SN~1006. 

The properties of image for the former possibility, namely ${\cal R}\rs{B}\neq 1$, are represented by 
\begin{equation}
 {\cal R}\rs{ic,Fermi}\simeq {\cal R}\rs{r} {\cal R}\rs{B}^{-(s+1)/2}
\end{equation}
instead of Eq.~(\ref{SN1006gmf:eq3_contrast}). So, having contrasts of brightness in Fermi image measured, we may directly have variations of MF strength over the SNR surface. Then this information may be used as input for the above sections providing us with measurements of the MF strength.

\section{Conclusions}
\label{sn1006gmf:conclusion_section}

H.E.S.S. cooperation has reported the detection of SN~1006 at the $1\%$ Crab flux 
level \citep{HESS-SN1006-2010}. In the SNR's image, two TeV lobes are visible coinciding with limbs 
in radio and X-rays \citep{HESS-SN1006-2010}. An application of the approach developed in \citet{Petetal09icp} to SN~1006 
provides argument that such coincidence may be considered as evidence about leptonic nature of TeV \g-ray emission of this SNR. 
The approach uses data of radio and X-ray observations as well as considers a number of assumptions about the strength and configuration of MF.

In the present note, that approach is ``inverted''. 
Namely, we propose a new, almost model-independent methods and analytical approximations, Eqs.~(\ref{SN1006cp:eq3}) and (\ref{SN1006gmf:eq3_contrast}), for estimation of the MF strength which uses contrasts in the radio, X-ray and TeV \g-ray images of SNR. The methods require however a \g-ray image with high spatial resolution and accuracy, comparable to the radio and X-ray maps. Unfortunately, HESS data prevent us from determination of the MF strength; present \g-ray data limit MF in limbs to values lower than few hundred $\mu$G. 
We believe that future observations considerably narrow the allowed range for MF strength.  

We show that the 
IC \g-ray limbs in SN~1006 should be in the same locations as in the radio and X-ray images for any $B\rs{o}>0.1\mu G$, if the 
aspect angle is $\leq 70^\mathrm{o}$ or the MF contrasts between NE and SE regions is $\leq 2.4$. 
If actual aspect angle or MF contrast are larger then some range of $B\rs{o}$ might be excluded by the fact of the same location of IC and synchrotron limbs.  This is true if the high-energy end of the electron spectrum is similar to a pure exponent (i.e. for $\alpha\approx 1$). In case it is broader ($\alpha<1$), the range of allowed aspect angles increases up to $90^\mathrm{o}$ or ${R}\rs{B}$ may be as large as 4 or even larger (as it could be, e.g., in case of the nonuniform ISMF). The narrower cut off ($\alpha>1$) may be excluded in SN~1006 for ${R}\rs{B}>2$ by the observed location of the bright TeV \g-ray limbs.

The number of unknown parameters in our methods (namely, two: ${\cal R}\rs{B}$ and $\alpha$) may be reduced with the use of the GeV image of SN~1006 which is expecting from the Fermi \g-ray telescope. Since electrons emitting GeV \g-rays have energies smaller than the cut off, this image is not sensitive to the variation of $E\rs{max}$ and, together with the radio image, may provide direct measurements of ${\cal R}\rs{B}$. If the contrast of brightness between the bright limbs on the Fermi image is the same as on the radio image, this provides a direct evidence about the leptonic nature of \g-rays in SN~1006.

We would like to note that our rough upper limit is higher than available estimates of the MF strength in SN~1006. 
Namely, \citet{Ber-Volk-2003-mf,Ber-et-al-2009} estimates the post-shock field from the width of the hard X-ray filaments as $\approx 150\un{\mu G}$ (this is strength deduced from the thinnest filament, therefore MF is expected to be smaller in other places). \citet{Vink2006} presents smaller estimate, $30\div 40\un{\mu G}$. The latter value is in agreement with the strength required by the `leptonic' model of the HESS collaboration \citep{HESS-SN1006-2010} where SNR is roughly considered as uniform plasma. More detailed calculations which consider distributions of MF and emitting particles inside SN~1006, yield $32\un{\mu G}$ for an average MF \citep[inverse-Compton scenario for TeV \g-rays,][]{sn1006cp} or $150\un{\mu G}$ for assumed uniform MF inside SNR 
\citep[scenario of dominating hadronic emission in TeV \g-rays,][]{Ber-et-al-2009}. 
\citet{sn1006cp} fits the same sharpest profile as used by \citet{Ber-Volk-2003-mf} with $95\un{\mu G}$ with a bit different model parameters. 
However, the post-shock MF in the limbs is estimated as $\leq 45\un{\mu G}$ since most of the radial profiles of the hard X-ray brightness are thicker than in the thinnest filament \citep{sn1006cp}.

\section*{Acknowledgments}
F.B. acknowledges partial support from the ASI-INAF agreement n. I/009/10/0.
O.P. acknowledges partial support from the program 'Kosmomikrofizyka' (NAS of Ukraine). 



\appendix
\section[]{Role of the possible concavity of the electron energy spectrum}
\label{sn1006gmf:appendix}

The volume emissivity of the population of relativistic electrons with the energy spectrum (\ref{ICpred:Espectrum}) is 
\begin{equation}
 q(\nu)=\int N(E)p(\nu,E)dE
\end{equation}
where $\nu$ is the frequency, $p$ is the single-electron emissivity. 

In case of the synchrotron process, $p\propto BF(\nu/\nu\rs{c})$, $\nu\rs{c}=c_1BE^2$, $F$ is the special function. 
It is known that in the range of electron energies where the spectrum (\ref{ICpred:Espectrum}) do not have abrupt changes, one can use the delta-function approximation of $F(\nu/\nu\rs{c})$ to obtain 
\begin{equation}
 q(\nu\rs{r})\propto KB^{(s+1-\delta s\rs{r})/2}
 \label{sn1006gmf:appendix:qr}
\end{equation}
for radio band and 
\begin{equation}
 q(\nu\rs{x})\propto KB^{(s+1-\delta s\rs{x})/2}\exp\left[-\left(\nu/\nu\rs{break}\right)^{\alpha/2}\right]
 \label{sn1006gmf:appendix:qx_delta}
\end{equation}
for X-rays; $\delta s\rs{r}$ and $\delta s\rs{x}$ corresponds to the electron energies which give maximum contribution to emission at radio and X-ray frequencies respectively. However, delta-function approximation is not adequate for the exponential part where the spectrum varies quite quickly; it is necessary to consider  the full single-particle emissivity there \citep{Reyn-98}. We need therefore to correct (\ref{sn1006gmf:appendix:qx_delta}) respectively. An approximation 
\begin{equation}
 q(\nu\rs{x})\propto KB^{(s+1-\delta s\rs{x})/2}\exp\left[-\beta\rs{x}\left(\nu/\nu\rs{break}\right)^{0.364\beta\rs{\alpha x}}\right]
 \label{sn1006gmf:appendix:qx}
\end{equation}
accurately restores the numerical convolution of the electron distribution (\ref{ICpred:Espectrum}) with the full single-particle
synchrotron emissivity (expressions for $\beta\rs{x}$ and $\beta\rs{\alpha x}$ are given in Sect.~\ref{sn1006cp:a_new_method_for_B}). 
Substitution of $K$ from (\ref{sn1006gmf:appendix:qr}) into (\ref{sn1006gmf:appendix:qx}) yields 
\begin{equation}
 q\rs{x}\propto q\rs{r}B^{-(\delta s\rs{x}-\delta s\rs{r})/2}
 \exp\left(-\beta\rs{x}\left(\frac{\nu\rs{x}}{\nu\rs{break}}\right)^{0.364\beta\rs{\alpha x}}\right).
 \label{ICpred:appendix:eq4}
\end{equation}
Comparison of this expression with Eq.~(\ref{ICpred:eq4}) (where $\delta s=0$) reveals the role of the eventual concave shape of the electron energy spectrum represented in Eq.~(\ref{ICpred:Espectrum}) by $\delta s(E)$.

In a similar fashion, the inverse-Compton emissivity of \g-rays with energy 1 TeV may be represented by the approximation
\begin{equation}
 q(\nu\rs{ic})\propto K\exp\left[-\beta\rs{ic}\beta\rs{\alpha ic}\left(c_1 B/\nu\rs{break}\right)^{0.375\beta\rs{\alpha x}}\right]
 \label{sn1006gmf:appendix:qgamma}
\end{equation}
(expressions for $\beta\rs{ic}$ and $\beta\rs{\alpha ic}$ are given in Sect.~\ref{sn1006cp:a_new_method_for_B}).
Substitution of $K$ from (\ref{sn1006gmf:appendix:qr}) into (\ref{sn1006gmf:appendix:qgamma}) yields (cf. with Eq.~\ref{ICpred:eq7})
\begin{equation}
 q\rs{ic}\propto q\rs{r} B^{-(s+1)/2}B^{\delta s\rs{r}/2}
 \exp\left(-\beta\rs{ic}\beta\rs{\alpha ic}\left(\frac{c\rs{1}B}{\nu\rs{break}}\right)^{0.375\beta\rs{\alpha x}}\right).
 \label{ICpred:appendix:eq7}
\end{equation}

It is clear now that $\delta s$ does not appear directly in the expressions for ratios of the IC \g-ray and synchrotron X-ray emissivities, ${\cal R}\rs{ic}=q\rs{ic1}/q\rs{ic2}$ and ${\cal R}\rs{x}=q\rs{x1}/q\rs{x2}$, of the two regions with the same MF strength $B$. Namely, the expressions are the same as (\ref{SN1006cp:eq1}) and (\ref{SN1006cp:eq2}). However, if the electron energy spectrum hardens toward $E\rs{max}$ then one has to use $s+\delta s\rs{x}$ and $s+\delta s\rs{ic}$ instead of $s$ in $\beta\rs{x}$ and $\beta\rs{ic}$. Nevertheless, $\xi$ in Eq.~(\ref{SN1006cp:eq3}) is almost insensitive to values of $s$ in the range $1.8\div 2.5$. Thus, possible hardening of the electron spectrum toward high energies does not affect results of the Sect.~\ref{sn1006cp:a_new_method_for_B}.

Let us consider the role of $\delta s$ in ratios ${\cal R}\rs{ic}$ and ${\cal R}\rs{x}$ for the two regions with different MF strengths, $B_1\neq B_2$. In this case, as it follows from (\ref{ICpred:appendix:eq7}), the right-hand side of Eq.~(\ref{SN1006gmf:eq3_contrast}) should be multiplied by ${\cal R}\rs{B}^{\delta s\rs{r}/2}$. Electrons emitting radio waves at 1.5 MHz in MF $B\sim 10\div 100\un{\mu G}$ have energies $\sim 100\un{MeV}$. The function $\delta s$ varies from 
0 at $E\simeq m\rs{e}c^2=0.5\un{MeV}$ to $\leq 0.5$ around $E\simeq E\rs{max}\sim 10\un{TeV}$. Around $E\sim 100\un{MeV}$, it reaches the value $\delta s\sim 0.1$. Therefore, ${\cal R}\rs{B}^{\delta s\rs{r}/2}\approx 1$ in wide range of ${\cal R}\rs{B}$. Thus, also in this case our results are negligibly affected by the eventual concavity of the electron energy spectrum. 

\label{lastpage}
\end{document}